\documentstyle[11pt,aspconf]{article}
\begin{document}
\input epsf.sty          

\title{Reverberation Mapping and Broad-Line Region Models}
\author{Dan Maoz}
\affil{School of Physics \& Astronomy and Wise Observatory,
 Tel-Aviv University, Tel-Aviv 69978, Israel}

\begin{abstract}
I review what we have learned about the BLR 
from reverberation mapping, point to some problems
and complications that have emerged, and outline some
future directions.
\end{abstract}


\section{Basics}
Reverberation mapping is a technique in which variability
data exhibiting light-travel-time effects are used to 
 to derive information on the geometry and kinematics
of a source. Reviews of the basic principles and methods of reverberation
mapping as applied to AGNs have appeared in Peterson (1988; 1993) and in
Netzer (1990). More detailed information
is compiled in the proceedings of a workshop dedicated
to the topic (Gondhalekar, Horne, \& Peterson 1994).

The fundamental equation of reverberation mapping
is the convolution equation relating the emission-line
 and continuum light curves, $L(t)$ and $C(t)$:
$$
L(t)=\int \Psi(\tau) C(t-\tau) d\tau.
$$
The kernel of the convolution is the transfer function, $\Psi(\tau)$,
which contains the geometrical and physical information inherent
to the reprocessing of energy that relates the two light curves.
It is the ``Green's function'' of the system, or the hypothetical
response of a BLR emission line to a very brief continuum burst.
An analogous two-dimensional
transfer function, $\Psi(\tau, v)$  relates
every projected-velocity bin, $v$, in the line profile to
the continuum light curve (see contribution by K. Horne).
$\Psi$ provides a ``picture''
of the BLR in time-delay (or time-delay / projected-velocity) space.
With some assumptions of symmetry,
one can then guess the full six-dimensional
geometry and kinematics of the BLR.

 In the next sections I will sketch my view of the current
status of the field, including how well the basic assumptions
of reverberation mapping seem to be faring, and what we
have learned to date about the BLR. I will outline
the directions I think echo mapping will take
in the near future.

\section{Problems}

With the execution of increasingly
ambitious reverberation-mapping campaigns (see 
this volume) and improved data, we
have seen a basic prediction of the AGN photoionization
model beautifully confirmed, namely,
 that emission-line light curves mimic the
continuum behavior, but with a lag due to light-travel time
effects. 
Among the most illuminating results are those from the joint {\it IUE}/optical
campaign on NGC 5548 in 1989
(Clavel et al. 1991; Peterson et al. 1991; Dietrich et al. 1993;
Maoz et al. 1993). The gross structure of the light curves
for the strong lines, whose sum is shown
and compared to the continuum in Figure 1, is just that expected.

 \begin{figure}
\epsfxsize=330pt
\epsfbox{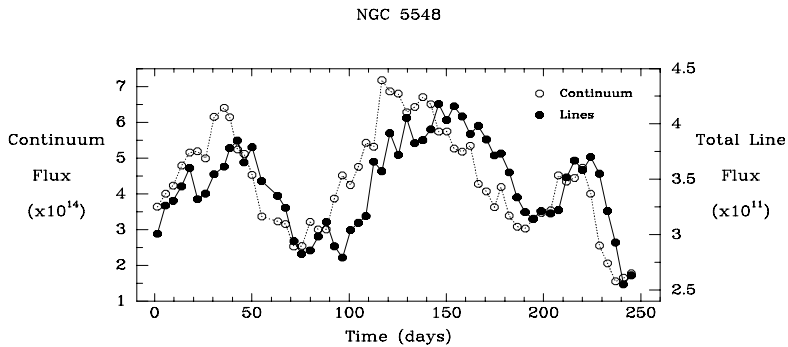} 
\caption{UV (1350\AA) continuum light curve (empty circles, left vertical
scale, in erg s$^{-1}$ cm$^{-2}$ \AA$^{-1}$) for NGC 5548,
 and total observed emission-line flux (filled circles,
right vertical scale, in erg s$^{-1}$ cm$^{-2}$ ),
 during the 1989 {\it IUE}/optical monitoring campaign.}
\end{figure}

However, the NGC 5548 data also
revealed various peculiarities, some of which point
to inadequacies in the basic assumptions of reverberation mapping.
Evidence for such effects has been seen in other objects as well.
The three main complications which seem to be indicated 
by variability data are non-linear response of the emission lines,
``misrepresentation'' of the ionizing continuum
by the observed optical or UV continuum,
and significant structural evolution of the BLR on timescales of a 
year. 
\begin{figure}
\epsfxsize=280pt
\epsfbox{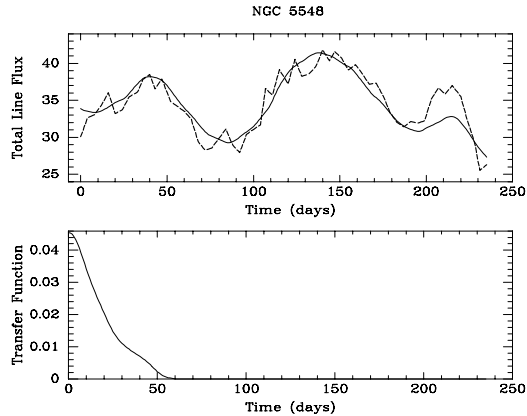} 
\caption{ Top panel: Jagged dashed line is the total emission line light
curve of NGC 5548, same as in Fig. 1. Bottom panel: Transfer function
obtained by maximum entropy inversion of the total light curve with
the UV continuum light curve, excluding the third ``event''. The smooth
solid line in the top panel is the reconstructed emission line light
curve obtained by convolving the continuum light curve with the transfer
function in the bottom panel. The third event is poorly
reproduced for any such monotonically decreasing transfer function.}
\end{figure}

The nonlinear behavior is most clearly visible in the NGC 5548
1989 light curve of CIV $\lambda 1549$. The total energy
in the third continuum ``event'' is much less than that of the 
previous two events and, if the line light curve were a linear
convolution of the continuum light curve, would produce a correspondingly
weak feature in the emission line light curve. Instead, the CIV
flux rises to the same amplitude it had in response to the previous
two events.
In retrospect, evidence for nonlinearities in BLR response has been around
for a long time. Wamsteker and Colina (1986) first pointed out
the saturation in the CIV $\lambda 1549$ level with increasing
continuum flux in Fairall-9. A more moderate manifestation of the same
effect, whereby CIV responds less than Ly$\alpha$ to continuum changes,
can be seen, e.g., in O'Brien et al. (1996) 
for the much smaller continuum variations of NGC 5548 in 1989.
In all AGNs the fractional amplitude of the emission
line variations is always considerably smaller than that of the continuum
variations, once known constant components such as narrow lines
and galaxy light have been accounted for. Clearly, some of the gas
contributing to the broad line flux does not respond to the {\it observed}
continuum variations. 

To a certain extent, non-linearities in the response of individual
lines are predicted by conventional photoionization models
(e.g. Goad, O'Brien \& Gondhalekar 1993).
However, the details of the response of a particular line cannot
be the entire story behind the problem, because as long as each BLR cloud
remains optically thick to the ionizing radiation, the energy
from each ionizing photon must come out as one line photon or 
another (or several). Therefore, the nonlinear aspects should
disappear in the total emission line light curve.(This is not
a new idea; see Blandford and McKee [1982], \S II.a.)
Returning to Fig.1, the third-event problem is, indeed,
 considerably reduced in the total line
light curve. However, the problem has not disappeared.

Figure 2 (top panel) shows the same total line curve (now shown
as a jagged dashed line). Superposed on it is a reconstructed
light curve (smooth solid line) obtained with a maximum entropy
inversion that was forced to produce a transfer function that
is monotonically decreasing (bottom panel). Maximum entropy
methods applied to the NGC 5548 data have generally produced
separate aliasing peaks in the transfer function which ``conspire''
with previous continuum events in order to ameliorate the
third-event problem. Figure 2 shows that if such aliasing is
not allowed, the amplitude of the third event is still much
too high, even in the total line flux.

Possible explanations are either that there is a fraction
of optically-thin gas in the BLR (Shields, Ferland, \& Peterson 1995;
O'Brien et al. 1996);
 or the continuum behavior we see is not the ionizing continuum
behavior that the BLR gas sees. This can come about if
the continuum emission is not emitted isotropically, or if 
the ionizing continuum is not strictly linearly proportional 
to the observed continuum longward of the Lyman edge
(as proposed, e.g., by Clavel \& Santos-Lleo 1990,
to explain the CIV saturation in Fairall-9).

There is already evidence for the latter possibility in NGC 5548,
in the well-established hardening of the continuum as it rises
(e.g. Maoz et al. 1993). Additional
evidence for continuum ``misbehavior'' comes from 
the intensive 1993 multiwavelength campaign on NGC 4151
(Crenshaw et al. 1996; Kaspi et al. 1996a; Warwick et al. 1996; 
Edelson et al. 1996). Figure 9 of Kaspi et al. (1996a)
 shows a scaled version of the
{\it IUE} 2700 \AA~ continuum light curve superposed on the 1275 \AA~
continuum light curve. The two light curves, just a factor of 2
apart in energy, are clearly not just linearly scaled versions
of each other. Things might be just as bad or worse
shortward of 912 \AA.
Additional evidence for problems with the ``surrogate'' continuum
is that
the optical continuum variation amplitude is similar
to that in a previous campaign on the same object (Maoz et al. 1991)
but the Balmer-line variations are much smaller in the
more recent campaign.
 If an imperfect correlation between the
observed and ionizing continua is the source of the
problems mentioned above, 
 it could a pose a difficult hurdle,
 as the ionizing continuum itself cannot be
directly observed.

A third complication is the detection of
a time-variable lag in several objects
(Netzer \& Maoz 1990; Peterson et al. 1994;
Wanders \& Horne 1994),
suggesting a BLR that evolves on timescales of about one year.
Possibly related is the fact that discrete velocity components
are known to appear and disappear in AGN line profiles, 
in a manner unrelated to the continuum variations
(Wanders and Peterson 1996). Wanders (this volume) has 
suggested that these components are fluctuations
in the fraction of orbiting BLR clouds that are illuminated by 
an ionization cone, rather than real changes
in the BLR geometry. In any case, the observed
evolution in the BLR provides an exciting new dimension 
in reverberation mapping, but may mean that long, sparsely-sampled
observations are not an alternative to intensive, season-long
campaigns.

Finally, there may have been
some
overinterpretation of the data.
It has become common to take variability
data and to attempt to invert the convolution equation directly to
recover $\Psi$ (e.g., Maoz et al. 1991; Krolik et al. 1991; Horne, Welsh
\& Peterson 1991; Peterson et al. 1994; Wanders \& Horne 1994).
However, even the best
currently available variability data are very noisy,
and therefore may produce non-unique 
transfer functions when deconvolved.
As an example, let us look at the H$\beta$ data for
NGC 5548. Horne et al. (1991) and Peterson et al. (1994)
have found that the maximum-entropy derived transfer 
function is peaked at 20 days, and has low-amplitude
at zero lag. They interpreted this as meaning that
there is little variable H$\beta$ line-emission coming 
from our line of sight to the nucleus, either because
of the BLR geometry or due to optical-depth effects
in the line. 

In Figure 3,  I
have taken the five-year-long optical continuum light
curve of NGC 5548 (Peterson et al. 1994; Korista et al. 1995),
linearly interpolated it to one-day intervals, and 
convolved it with three different transfer functions:
a delta-function peaked at 20 days, a top-hat function
that is positive from 0 to 40 days, and a triangular 
function peaked at 0 days and decreasing to zero at
60 days. In my talk, I argued that
the differences between the light curves
produced by these very different-shaped transfer functions are
minute compared to the uncertainties in the line measurements
themselves, and that these data
cannot distinguish between these transfer functions, and
in particular between transfer functions peaked at zero-lag
and away from zero. After my talk, Keith Horne challenged me
to add simulated measuring errors to the fake light curves,
and send them to him, which I did. To my surprise, Keith's
MEMECHO program recovered the transfer functions well enough
for him to easily guess by a process of elimination the correct
transfer function to assign to each light curve. On the other hand,
 Wanders and Peterson (1996) have recently
concluded from analysis of a newly-reduced
version of the same observations that the NGC 5548 H$\beta$ transfer 
function cannot yet be uniquely determined.


\begin{figure}
\epsfxsize=250pt
\epsfbox{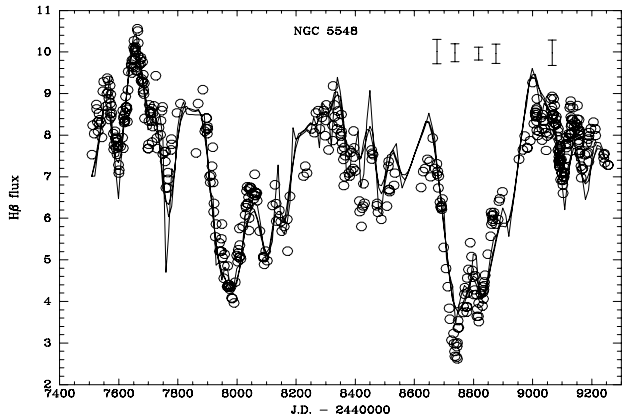} 
\caption{NGC 5548 5-year H$\beta$ light curve (empty
circles). Some typical error bars are shown in upper right
corner. The three solid lines are the convolution of the interpolated
optical continuum light curves with
three different-shaped transfer functions: a delta-function,
a top-hat, and a triangle peaked at zero lag. Can the data distinguish
between these models?}
\end{figure}

\section{What Have We Learned?}

Despite the problems outlined above,
we have learned quite a bit about the BLR from reverberation
mapping. 
An important result from the 1989 NGC 5548 capaign,
 supported by the 1993 {\it IUE}+{\it HST} campaign
of the same object (Korista et al. 1995),
 is that the BLR is stratified in 
ionization. Different emission lines respond to the
continuum with different lags. The range in lags spans
a factor of 5, perhaps more, and there is a trend
for the most highly ionized species to have the
smallest lags (i.e., be emitted at preferentially small
radii). This result has outmoded the single-cloud
photoionization models that were common for many years.
It is now clear that BLR gas exists at a range of radii.
Baldwin et al. (1995; see
also Baldwin, this volume) have argued that such
stratification is a natural
consequence of a ``thick'' BLR, with different lines
reaching their peak emission efficiencies at different
radii.

The question of BLR geometry and kinematics is still
unresolved. I note that the data on
the best-studied object, NGC 5548, have produced 
papers advocating a variety of models. Wanders et al. (1995)
argue for a spherical BLR with randomly inclined 
circular orbits, illuminated by an ionizing bi-cone
viewed approximately end-on. (This geometry
is probably indistinguishable from one with random
radial orbits, with the bi-cone approximately in the plane
of the sky.) Done \& Krolik (1996) find the data consistent
with a thick spherical geometry with Keplerian orbits.
Chiang \& Murray (1996; see also Murray, this volume) and Rokaki (this volume)
 model the results with a Keplerian disk.
The data cannot yet 
distinguish between these models. We probably {\it can}
say that the BLR transfer function is resolved on
timescales of days, and therefore has some structure.
 I hope that future experiments will see the details
of this structure.

We do know
some things that the BLR is {\it not}. Models invoking
pure radial flow (whether infall or outflow)
as the dominant line broadening mechanism have been long discussed.
A prediction of any such model is a lag between the
blue and red wings of the profile, comparable to the lag
of the total line flux behind the continuum. Every experiment
that has tested this prediction with adequate temporal resolution
has obtained a null result (e.g. Maoz et al. 1991, Wanders et al. 1995).
It will be interesting to see if these null results hold
up in higher luminosity objects; radial flows
are utilized to explain the line shifts observed in
quasars (see Espey, this volume).

\section{The BLR at Higher Luminosities}

Until recently, quasars have been largely neglected by echo mappers.
 This has been due to their faintness (and hence 
inaccesibility to {\it IUE} and to small telescopes), their frequent
lack of narrow emission lines (still the most popular flux 
calibrator), and some prejudices about their (presumably long
and non-paper/thesis-producing) variation and response timescales.
The small amount of data that did exist on quasar emission-line
variability (often only two or three epochs per object)
produced controversial and sometimes contradicting
interpretations (see Peterson 1993, for a review).

Happily, reverberation results for higher-luminosity AGNs
have begun to appear. Carone et al. (1996) measured a surprisingly
large H$\beta$/continuum lag (100 days) in Mrk 509, a luminous Seyfert
galaxy.
For the past five years,
a collaboration between Wise and Steward Observatories has been
monitoring 28 bina-fide quasars from the PG sample (Maoz et al. 1994;
Kaspi et al. 1996b; see Kaspi, this volume). This program has
demonstrated that reverberation mapping indeed works in quasars,
much as it does in Seyferts. The first clear lags that have been
measured, in two quasars, are also of about 100 days.

Does the BLR size scale with 
luminosity as $R_{BLR}\propto L^{1/2}$,
as long predicted based on
the overall great similarity of AGN spectra over many orders of
magnitude in luminosity?
Figure 3 in Kaspi's contribution (this volume)
 shows the BLR size as a function
of luminosity. The two highest luminosity points are the two
new quasar measurements.
While more quasar BLR radii (which are upcoming
from the Wise-Steward program) are desirable,
the trend in the figure is certainly suggestive
of the expected relation. 

\section{The Future}
With the demise of {\it IUE} (after an incredibly long and fruitful service)
a rethinking of strategy is required of echo mappers. To continue progress
in understanding the BLR of Seyferts by this technique, a significant
improvement is required in the sampling frequency (approximately
daily sampling lasting about 6 months) and in the number of AGNs
that are monitored. High spectral resolution and S/N may also start 
producing the long-sought 2-D transfer functions. Measurement errors
should remain below the 2-3\% level, since we now know that relatively
low-amplitude variations (tens of \%) are the norm in non-blazar AGNs.
All these constraints point to the need for one or more dedicated 
AGN-monitoring ground-based telescopes. High efficiencies could be
obtained with 2-3m class apertures, especially if a high degree
of automation is implemented. Two telescopes can 
minimize to a large extent the gaps due to weather.

Progress with quasars may be easier. Going down to $B=20$ mag, there
are 15 quasars per square degree. Using multi-object spectrographs
with wide fields, such as are available on many large telescopes,
one can monitor tens of quasars simultaneously. The slower response
time (further dilated by $1+z$ in high-$z$ objects) also means
that the sampling constraint can be relaxed. With once-a-week
observations on a 4m telescope over several years, the BLRs
of hundreds of quasars, spanning both luminosity and look-back time,
can be studied.

\section{Conclusions}

The work of the past decade has demonstrated that
reverberation mapping works
in both low and high luminosity AGNs, and
has huge potential. While we cannot yet say that we have
determined the structure or kinematics of any BLR, we
are at the point where we can discern time-resolved transfer
functions in the best-studied objects. Improved data 
will reveal the details of the transfer functions, and perhaps
of the BLRs themselves. We have already learned about 
ionization stratification, and the inapplicability of some
models, e.g. pure radial flow models for Seyferts.

The steadily improving quality of the observations have
also revealed complications, indicative of
 the approximations behind some of the
basic assumptions: non-linear response, suggestive of the
presence of some optically-thin gas, and BLR evolution,
vs. the usually assumed stationarity. These are ``good''
complications, in that they point us to a more realistic
picture of the BLR. A more worrisome complication is the possibility
of a ``misbehaving'' observed continuum that does not
completely represent the
ionizing continuum seen by the BLR.
Future, efficient, reverberation surveys, such as I have outlined,
could show the degree to which such continuum misbehavior is a problem.




\begin{references}

\reference Baldwin, J., Ferland, G., Korista, K., \& Verner, D. 1995, 
ApJ, 455, L119

\reference Blandford, R.D. \& McKee, C.F. 1982, \apj, 255, 419

\reference Carone, T.E., et al. 1996, ApJ, in press

\reference Chiang, J., \& Murray, N. 1996, ApJ, in press

\reference Clavel, J., et al. 1991, \apj,  366, 64

\reference Clavel, J. \& Santos-Lleo, M. 1990, A\&A, 230, 3 

\reference Crenshaw, D. M. et al. 1996, ApJ, 470, in press

\reference Dietrich, M. et al. 1993, ApJ, 408, 416

\reference Done, C. \& Krolik, J.H. 1996, ApJ, in press

\reference Edelson, R.A., et al.  1996, ApJ, 470, in press

\reference Goad, M.R., O'Brien, P.T., \& Gondhalekar, P.M. 1993, MNRAS, 263, 149


\reference Gondhalekar, P.M, Horne, K., \& Peterson, B.M. 1994, Reverberation 
Mapping of the Broad Line Region in Active Galactic Nuclei, San Francisco: ASP

\reference Horne, K., Welsh, W.F., \& Peterson, B.M. 1991, \apj, 367, L5

\reference Kaspi, S., et al. 1996a, ApJ, 470, in press

\reference Kaspi, S., Smith, P.S., Maoz, D., Netzer, H., \& Jannuzi, B.T. 1996b,
ApJL, in press

\reference Korista, K.T., et al. 1995, ApJS, 97, 285

\reference Krolik, J.H., Horne, K., Kallman, T.R., Malkan, M.A.,
Edelson, R.A., \& Kriss, G.A. 1991, \apj, 371, 541




\reference Maoz, D.,
 Netzer, H., Leibowitz, E., Brosch, N., Laor, A., 
Mendelson, H., Beck, S., Almoznino, E., and Mazeh, T.
 1991, \apj,  367, 493

\reference Maoz, D., et al. 1993, \apj, 404, 576

\reference Maoz, D., Smith, P.S., Jannuzi, B.T., Kaspi, S., \& Netzer, H.
1994,\apj, 421, 34

\reference Netzer, H. 1990, in Saas-Fee Advanced Course 20, ed. R.D.
Blandford, H. Netzer, \& L. Woltjer, Berlin: Springer-Verlag

\reference Netzer, H. \& Maoz, D. 1990, \apj, 365, L5


\reference O'Brien, P.T., Goad, M.R., \& Gondhalekar, P.M. 1996, MNRAS, in press

\reference Peterson, B.M., 1988, PASP, 100, 18


\reference Peterson, B.M., et al. 1991, \apj,  368, 119


 \reference Peterson, B.M. 1993,  PASP,  105, 247

\reference Peterson, B.M., et al. 1994,\apj, 425, 622

\reference Shields, J.C., Ferland, G.J., \& Peterson, B.M. 1995, ApJ, 441, 507






\reference Wamsteker, W. \& Colina, L. 1986, ApJ, 311, 617

\reference  Wanders, I., et. al., 1993, A\&A, 269, 39

\reference Wanders, I., \& Horne, K. 1994, A\&A, 289, 76


\reference Wanders, I. et al. 1995, ApJ, 453, L87

\reference Wanders, I. \& Peterson, B.M. 1996, ApJ, in press

\reference Warwick, B., et al. 1996, ApJ, 470, in press
\end{references}
\end{document}